\documentclass[journal]{IEEEtran}

\makeatletter
\long\def\@makecaption#1#2{\ifx\@captype\@IEEEtablestring%
\footnotesize\begin{center}{\normalfont\footnotesize #1}\\
{\normalfont\footnotesize\scshape #2}\end{center}%
\@IEEEtablecaptionsepspace
\else
\@IEEEfigurecaptionsepspace
\setbox\@tempboxa\hbox{\normalfont\footnotesize {#1.}~~ #2}%
\ifdim \wd\@tempboxa >\hsize%
\setbox\@tempboxa\hbox{\normalfont\footnotesize {#1.}~~ }%
\parbox[t]{\hsize}{\normalfont\footnotesize \noindent\unhbox\@tempboxa#2}%
\else
\hbox to\hsize{\normalfont\footnotesize\hfil\box\@tempboxa\hfil}\fi\fi}
\makeatother

\ifCLASSINFOpdf
\else
   \usepackage[dvips]{graphicx}
\fi
\usepackage{url}
\usepackage{graphicx}
\usepackage{amsmath,graphicx}
\usepackage{cite}
\usepackage{scrextend}
\usepackage[symbol]{footmisc}
\usepackage{xcolor}
\usepackage{soul}

\usepackage{multirow}

\begin{document}

\title{Adaptive Debanding Filter}

\author{Zhengzhong Tu, Jessie Lin, Yilin Wang, Balu Adsumilli, and Alan C. Bovik, \IEEEmembership{Fellow, IEEE}
\thanks{Z. Tu and A. C. Bovik are with Department of Electrical and Computer Engineering, The University of Texas at Austin, Austin, TX, 78712, USA (email: zhengzhong.tu@utexas.edu, bovik@utexas.edu).}
\thanks{J. Lin, Y. Wang and B. Adsumilli are with Google Inc., Mountain View, CA, 94043, USA. (email: jessielin@google.com, yilin@google.com, 
badsumilli@google.com)}
}

\markboth{}
{Shell \MakeLowercase{\textit{et al.}}: Bare Demo of IEEEtran.cls for IEEE Journals}
\maketitle

\begin{abstract}
Banding artifacts, which manifest as staircase-like color bands on pictures or video frames, is a common distortion caused by compression of low-textured smooth regions. These false contours can be very noticeable even on high-quality videos, especially when displayed on high-definition screens. Yet, relatively little attention has been applied to this problem. Here we consider banding artifact removal as a visual enhancement problem, and accordingly, we solve it by applying a form of content-adaptive smoothing filtering followed by dithered quantization, as a post-processing module. The proposed debanding filter is able to adaptively smooth banded regions while preserving image edges and details, yielding perceptually enhanced gradient rendering with limited bit-depths. Experimental results show that our proposed debanding filter outperforms state-of-the-art false contour removing algorithms both visually and quantitatively.
\end{abstract}


\IEEEpeerreviewmaketitle

\section{Introduction}

Recent years have witnessed significant advancements of video compression technologies. The implementation of modern video coding standards, such as H.264/AVC \cite{wiegand2003overview}, HEVC \cite{sullivan2012overview}, VP9 \cite{mukherjee2013latest}, and AV1 \cite{chen2018overview} have greatly benefited high-quality video streaming applications over bandwidth-constrained networks. Despite the improved capabilities of video codecs, banding artifacts remain  a dominant visual impairment of high-quality, high-definition compressed videos. Banding artifacts, which present as sharply defined color bands on otherwise smoothly-varying regions, are often quite noticeable, in part because of the tendency of visual apparatus to enhance sharp gradients, as exemplified by the Mach bands illusion \cite{ratliff1965mach}. To further optimize the perceptual quality of compressed user-generated videos \cite{tu2020ugc, tu2020comparative}, developing ways to detect and reduce banding artifacts is a problem of pressing interest.

\begin{figure}[!t]
\centering
\def\xlinewidth{0.16}
\def\hswidth{-0.4em}
\def\xem{2pt}
\footnotesize
\setlength{\tabcolsep}{2pt}
\begin{tabular}{ccc}
  \multicolumn{3}{c}{\includegraphics[ width=0.35\linewidth, keepaspectratio]{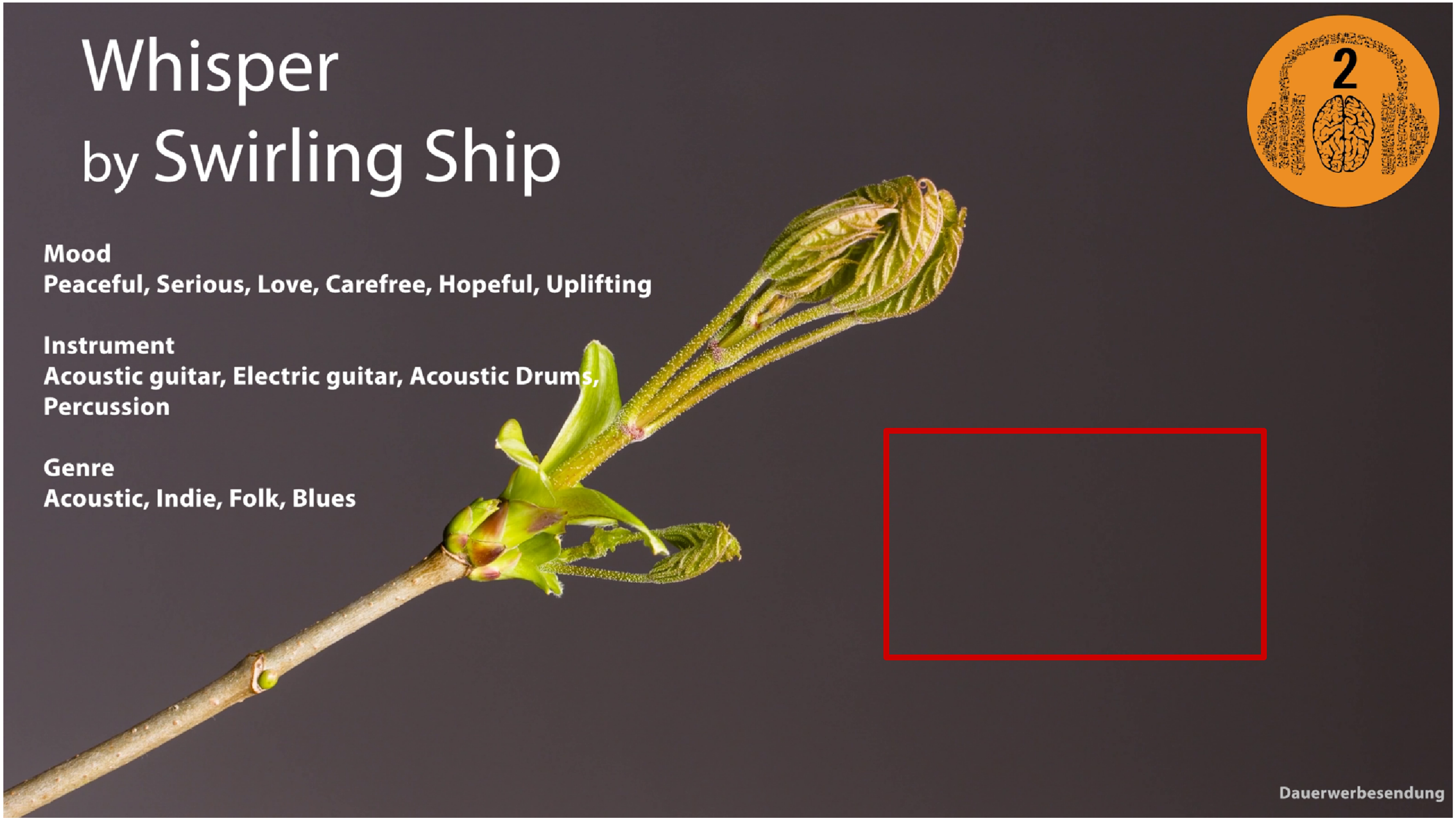}}
 \\[\xem]
  \includegraphics[ height=\xlinewidth\linewidth, keepaspectratio]{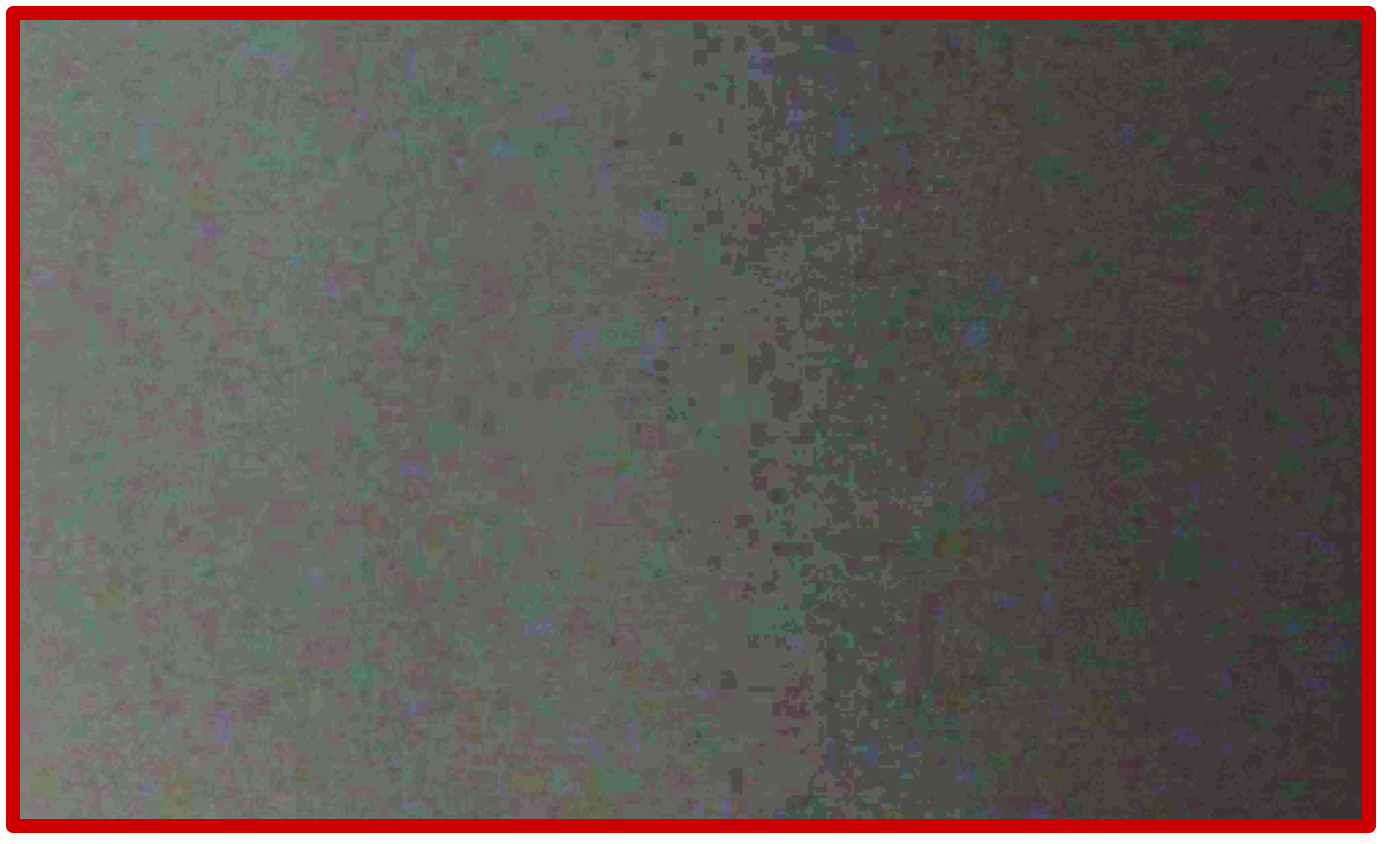} &
  \includegraphics[ height=\xlinewidth\linewidth, keepaspectratio]{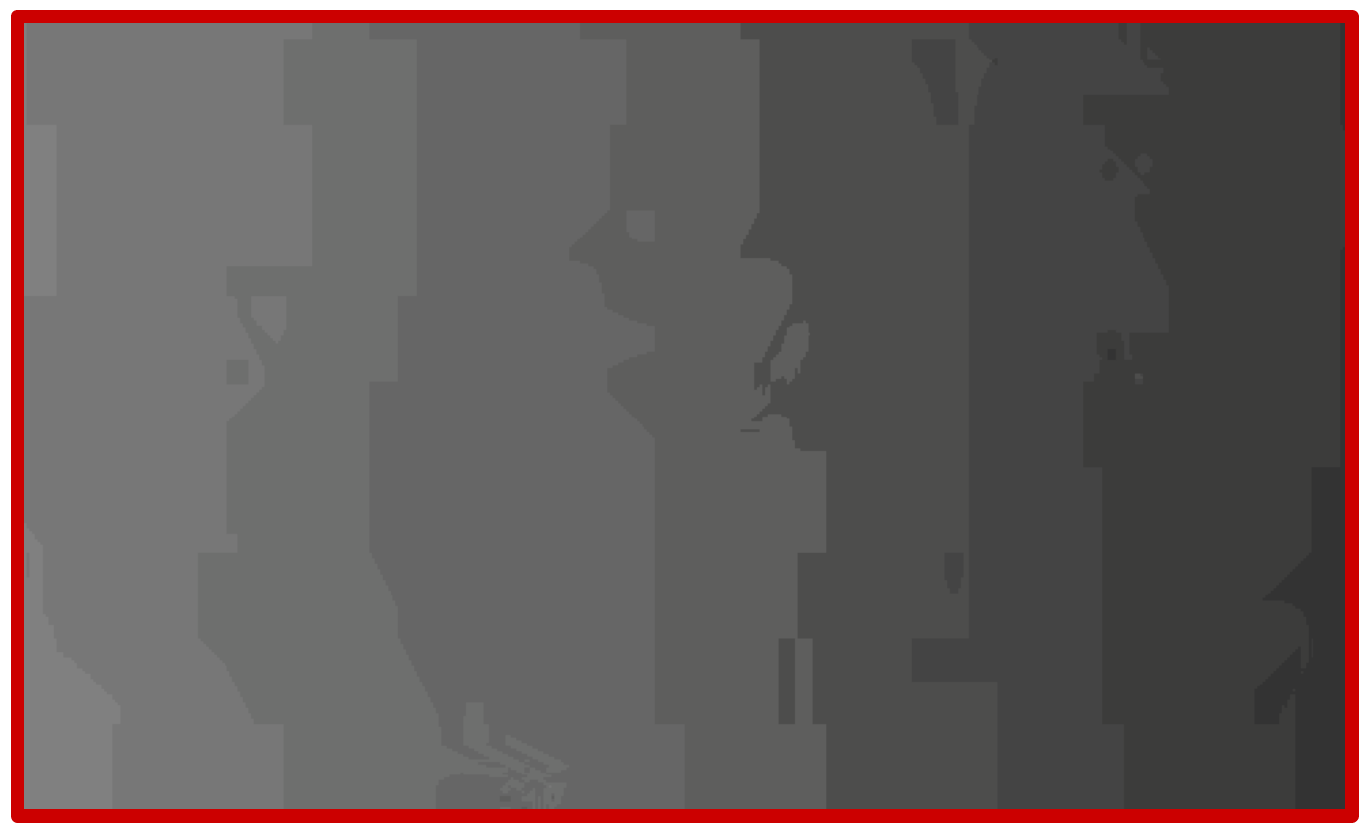} &
  \includegraphics[ height=\xlinewidth\linewidth, keepaspectratio]{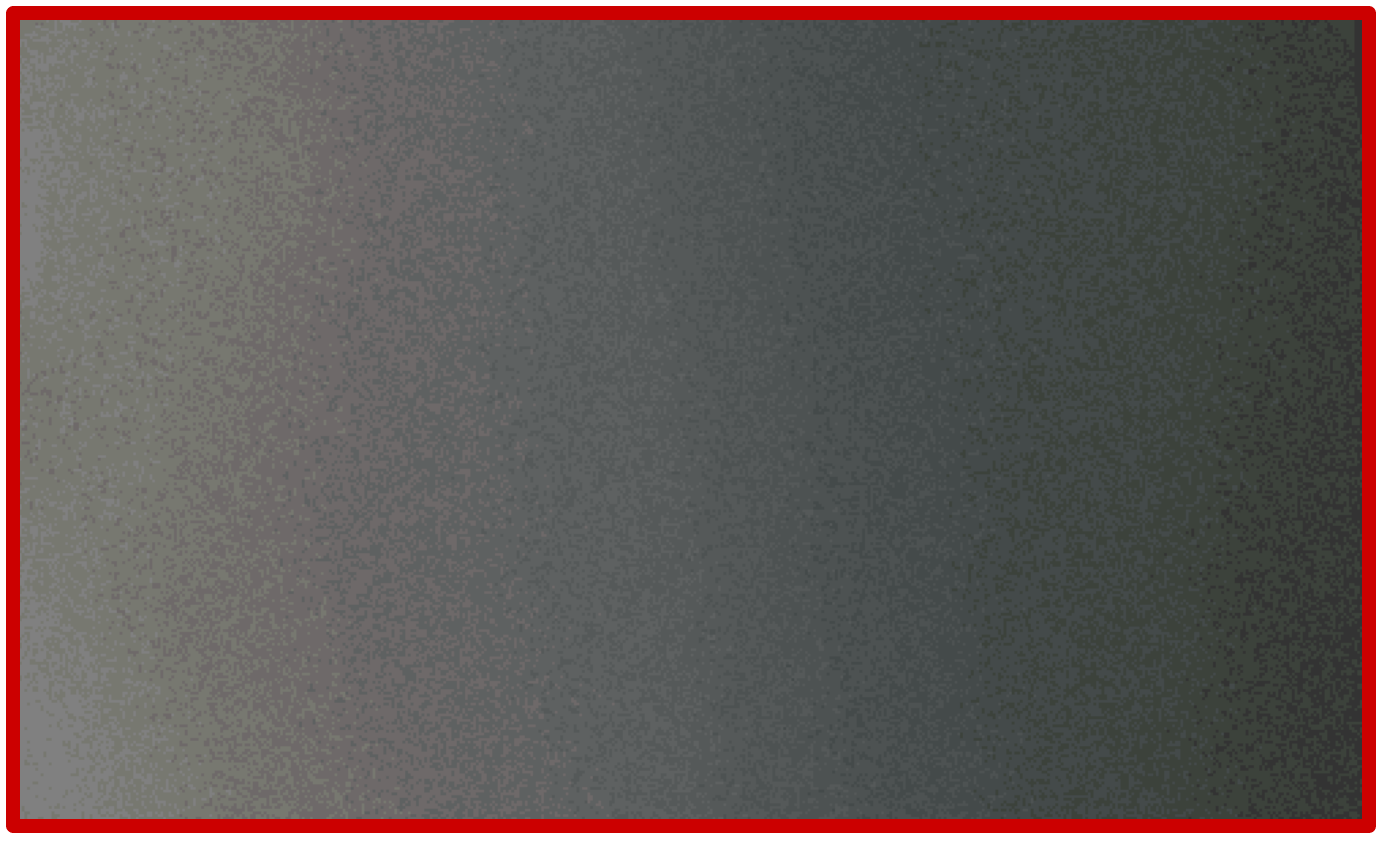}  \\[\xem]
  (a) Original &
  (b) Compressed &
  (c) Debanded \\
\end{tabular}
\caption{Examples of banding artifacts introduced by video encoding/transcoding: the first row shows an example of banded videos, while the second row shows contrast-enhanced zoom-in of the box-marked banding area in (a) original, and (b) VP9-compressed video, respectively. Our proposed AdaDeband filter effectively processes the banded regions, yielding perceptually enhanced smoothness rendering, as shown in (c).}
\label{fig:example_deband}
\end{figure}

\begin{figure*}[!t]
\centering
\includegraphics[width=0.86\textwidth]{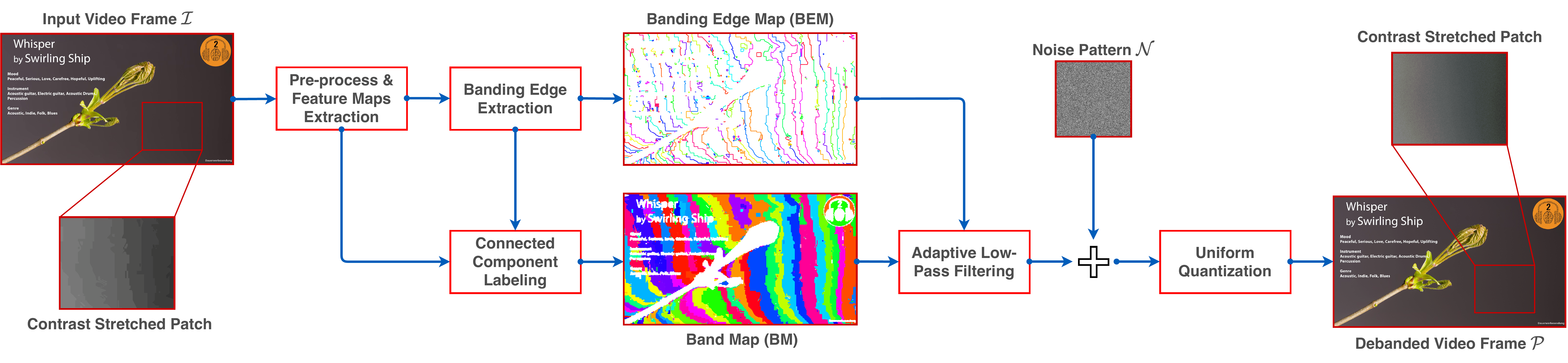}
\caption{Schematic overview of the proposed adaptive debanding filter (AdaDeband).}
\label{fig:flowchart}
\end{figure*}

Prior approaches to banding reduction may be categorized in three ways. If debanding is attempted on the source content before encoding, it is a pre-processing step \cite{roberts1962picture, joy1996reducing, daly2004decontouring, daly2003bit}. However, these methods have generally been developed for decontouring heavily quantized pictures rather than on compressed/banded video content. A second approach is to apply in-loop processing, whereby the quantization process is adjusted inside the encoder to reduce banding effects \cite{yoo2009loop, casali2015adaptive}. The third approach, post-filtering, has been most widely studied, since it offers maximum freedom for decoder implementations, i.e., the design of post-filters can be relatively unconstrained and flexible. Most banding removal algorithms follow a two-step procedure: first, banding regions are detected and located in the source video frame; then, spatially local filtering is applied to reduce the banding artifacts with dithering sometimes incorporated. For the banding detection stage, some methods \cite{daly2004decontouring,ahn2005flat,choi2006false, lee2006two,huang2016understanding,bhagavathy2009multiscale} exploit local features, either pixel- or block-wise, such as the image gradient, contrast, or entropy, to measure potential banding statistics. Other methods utilize image segmentation techniques \cite{wang2016perceptual, baugh2014advanced}. Either way, banding artifacts are subsequently suppressed by applying low-pass smoothing filters \cite{daly2004decontouring, lee2006two, choi2006false}, dithering techniques \cite{wang2014multi, bhagavathy2009multiscale, jin2011composite, ahn2005flat}, or combinations of these \cite{baugh2014advanced, huang2016understanding}.


We deem post-filtering a better approach to handle banding artifacts, since it can be performed outside of the codec loop, e.g., in the display buffer, hence offering maximum freedom of design. Moreover, state-of-the-art adaptive loop filters like those implemented in VP9 or HEVC have not been observed to supress or have any effect on banding artifacts \cite{wang2016perceptual, huang2016understanding}. Here we propose a new adaptive debanding filter, which we dub the AdaDeband method, as a post-processing solution to reduce perceived banding artifacts in compressed videos. Unlike many prior debanding/decontouring models, we recast the problem differently, as a reconstruction-requantization problem, where we first employ an adaptive interpolation filter along each banded region to estimate the ``ideal'' plane at a higher bit-depth, then we re-quantize the signal to 8-bit via a dithering technique to suppress quantization error, yielding a visually pleasant enhancement of the banding region. Fig. \ref{fig:example_deband} exemplifies VP9-compressed banding artifacts and the corresponding AdaDeband-filtered output. We demonstrate that our proposed method outperforms other recent debanding methods, both qualitatively and quantitatively. 

The rest of the paper is structured as follows. We present the details of the proposed adaptive debanding filter in Section \ref{sec:deband-filter}, while evaluative results are given in Section \ref{sec:exp}. Finally, Section \ref{sec:conc} concludes the paper.

\section{Adaptive Debanding Filter}
\label{sec:deband-filter}

Fig. \ref{fig:flowchart} illustrates a schematic of AdaDeband, which is comprised of three modules. First, a banding detector is deployed to localize the banding edges and band segments with pixel precision. A content-aware size-varying smoothing filter is then applied to reconstruct the gradients within band segments at a higher bit-depth, while preserving image texture details. The final step involves re-quantizing the reconstructed smooth regions to SDR 8-bit resolution using a dithering process to reduce quantization error, yielding a perceptually pleasant debanded image.

\subsection{Banding Region Detection}
\label{ssec:band-detect}

We utilize the Canny-inspired blind banding detector with same parameters proposed in \cite{tu2020bband} for banding edge extraction. Only the first two modules, pre-processing and banding edge extraction, are performed to obtain a banding edge map (BEM). For convenience, we restate some definitions from \cite{tu2020bband}: after self-guided pre-filtering, pixels having gradient magnitudes less than $T_1$ are labelled as flat pixels (FP); pixels with gradient magnitudes exceeding $T_2$ are marked as textured pixels (TP). The remaining pixels are grouped into a candidate banding pixel (CBP) set, based on which the BEM is extracted. Connected-component (CC) labeling is applied on the set of non-textured pixels (FP $\cup$ CBP), thereby generating a band map (BM). Band edges (BE) define the boundaries of adjacent bands (B); i.e., the bands (B) are framed by band edges (BE), as shown in Fig. \ref{fig:filter}. In this way we define and extract two component sets, BEM and BM, which together compose all the banded regions of a given video frame.

\subsection{Banding Segment Reconstruction}

\begin{figure}[!t]
\centering
\footnotesize
\def\imghei{0.20\textwidth}
\begin{tabular}{cc}
\includegraphics[height=\imghei]{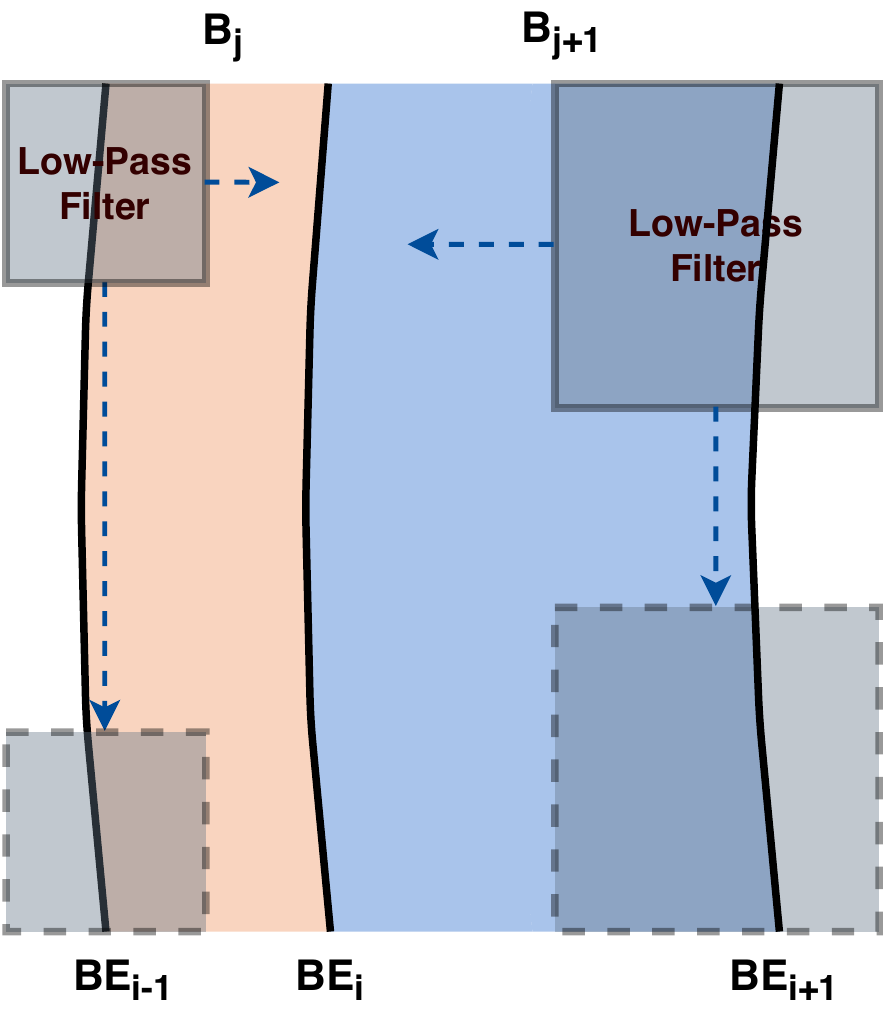}  & 
\includegraphics[height=\imghei]{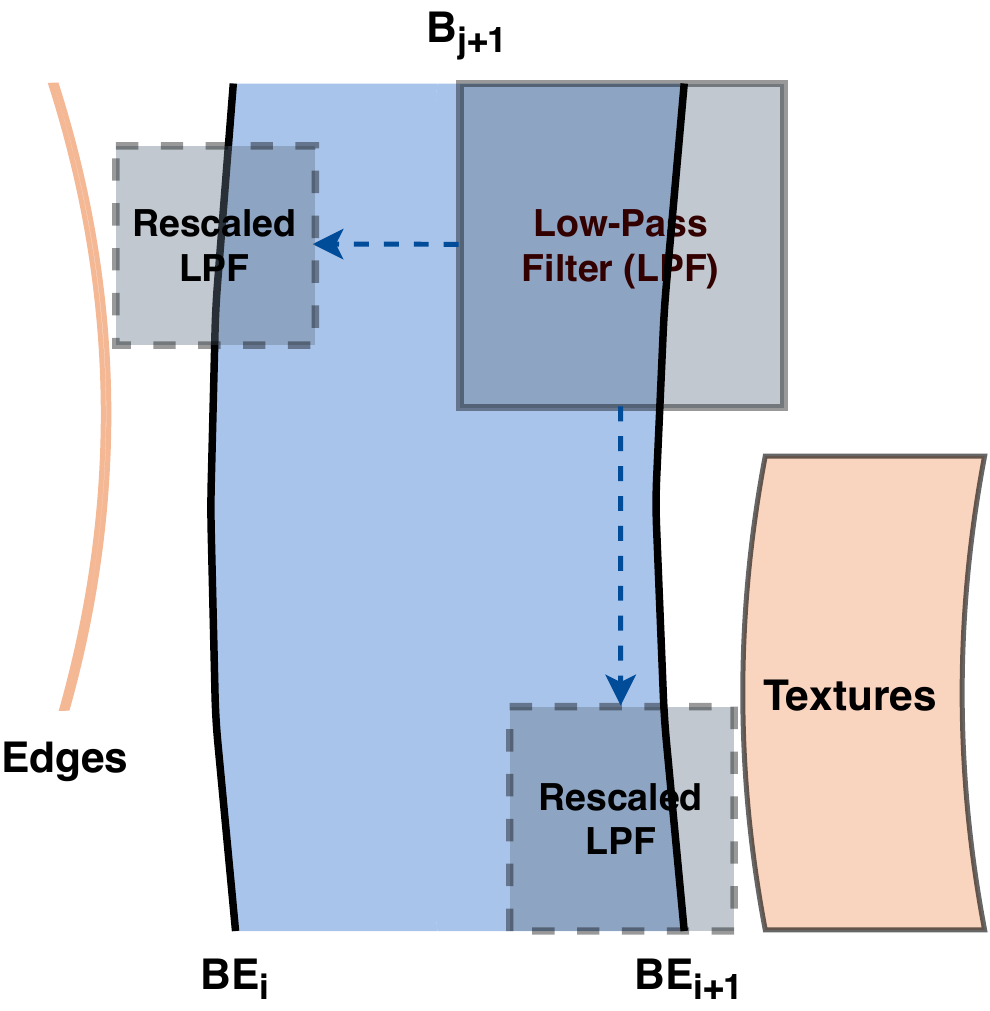}  \\ 
(a) Size variability & 
(b) Content (Texture) awareness \\
\end{tabular}
\caption{The size of the low-pass filter window that is applied on different bands depends on both (a) the band widths and (b) contextual edges/textures.}
\label{fig:filter}
\end{figure}

Banding artifacts usually occur on regions of small (but non-zeros) gradients, usually appearing as bands separated by steps of color and/or luminance. As depicted in Fig. \ref{fig:filter}, pixels lying within a band have very similar colors and luminances. A simple method of restoring an `original' smooth local image is to apply a low-pass filter (LPF) to interpolate across bands. A traditional 2D-LPF may be formulated as:
\begin{equation}
\label{eq:lpf}
\mathcal{J}(i,j)=\sum_{k=-K}^K\sum_{\ell=-L}^Lw_{k,\ell}\mathcal{I}(i-k,j-\ell),
\end{equation}
where $\mathcal{I}(i,j)$ are input pixel (luminance or color) values at spatial locations $(i,j)$, and $w_{k,\ell}$ are the filter coefficients, where $\sum_{k,\ell}w_{k,\ell}=1$. Here we use the simple moving average filter ($w_{k,\ell}=1/[(2K+1)(2L+1)]$) as the smoothing LPF, although one may use any other smoothing filter, such as a Gaussian-shaped filter, or even a nonlinear device such as a median filter.  

To effectively smooth band discontinuities, the span of LPF should be adequately wide relative to the band width. In Fig. \ref{fig:filter}(a), for example, it is only necessary to apply a small LPF on the pink band, whereas a larger LPF would be required for the blue segment. Thus, we apply a dynamic way of determining the filter size for each band, or segment of a band, if the width of the band varies along its axis, based on the extracted BEM and BM. Assume B\textsubscript{j} to be an exemplar band, which is framed by detected band edges, BE\textsubscript{i-1} and BE\textsubscript{i}, as depicted in Fig. \ref{fig:filter}(a). For all the pixels $\mathcal{I}(m,n)$ located within band B\textsubscript{j}, the spatial extent of the LPF is defined in terms of the ratios of the band area to the lengths of the adjacent BEs. In the unusual instance where a band is enclosed by a single BE, then the ratio is scaled by four. Both cases are expressed here: 
\begin{equation}
\label{eq:l}
l(m,n)=
\begin{cases}
4\times|\mathrm{B}_j|/|\mathrm{BE}_k|,\ k\in\mathcal{N}_{\mathrm{B}_j} & if\ |\mathcal{N}_{\mathrm{B}_j}|=1 \\
\max\limits_{k}|\mathrm{B}_j|/|\mathrm{BE}_k|,\ k\in\mathcal{N}_{\mathrm{B}_j} & if\ |\mathcal{N}_{\mathrm{B}_j}|>1
\end{cases}
,
\end{equation}
where $|S|$ is the cardinality of the pixel set $S$ (area of band, or length of band edge), and $\mathcal{N}_{\mathrm{B}_j}$ denotes the set of band edges that enclose B\textsubscript{j}. Finally, define the space-varying radius of the LPF window at $(m,n)$ to be half of $l(m,n)$:
\begin{equation}
\label{eq:h}
h(m,n)=\max\{1,\lfloor(l(m,n)-1)/2)\rfloor\}.
\end{equation}

We have ensured that the span of the LPF adapts to the local geometries of banded areas, as shown in Fig. \ref{fig:filter}(a). Nevertheless, it must also be recognized that the LPFs may process visually important textures or object boundaries while smoothing large banding regions. This kind of content blurring is not acceptable. Accordingly, we constrain the LPF not to include any TP in the sampling window, which is achieved by recursively halving the filter size (as in Fig. \ref{fig:filter}(b)):
\begin{equation}
\label{eq:halve}
\begin{split}
\textbf{repeat} & \ h(m,n):=\max\{1,\lfloor h(m,n)/2\rfloor\}\\ 
\textbf{until} & \ \forall\ (i,j)\in \mathcal{H}^h(m,n), (i,j)\notin \mathrm{TP}
\end{split}
\end{equation}
where $\mathcal{H}^h(m,n)=\{(m+x,n+y)\},\ x=-h,...,h,\ y=-h,...,h,$ is the set of indices of the (square) filter window centered at $(m,n)$, with linear dimensions $2h+1$. 

The above filter-size-determining process is performed on every banded pixel in set CBP to generate a window-size map for further deployment of size-varying LPFs. We observed that, however, the estimated window-size map tends to be noisy due to the potential unrobustness of banding detector. Thus, a median filter is applied to further ``denoise'' the estimated window-size map, based on which LPFs are then conducted via Eq. (\ref{eq:lpf}).

\subsection{Requantization with Dithering}
\label{sec:dither}

Dithering techniques are used in a variety of ways in the design of debanding algorithms. Some methods, for example, only apply filtering without dithering to remove false contours \cite{daly2004decontouring, lee2006two, choi2006false}. Other methods apply only a very small amount of dither, either by adding random noise (noise-shaping) \cite{yoo2009loop, bhagavathy2009multiscale, jin2011composite, wang2014multi}, or by stochastic shuffling of pixels \cite{ahn2005flat, huang2016understanding}, but without any smoothing filter. Among those that combine filtering with dithering, Baugh \cite{baugh2014advanced} proposed to apply dithering after smoothing on banded regions, while the authors of \cite{huang2016understanding} suggested probabilistic dithering prior to average filtering. Here we will show that dithered re-quantization very effectively ameliorates banding artifacts arising from compression.



\begin{figure}[!t]
\centering
\footnotesize
\def\imghei{0.18\textwidth}
\begin{tabular}{cc}

 \includegraphics[width=\imghei]{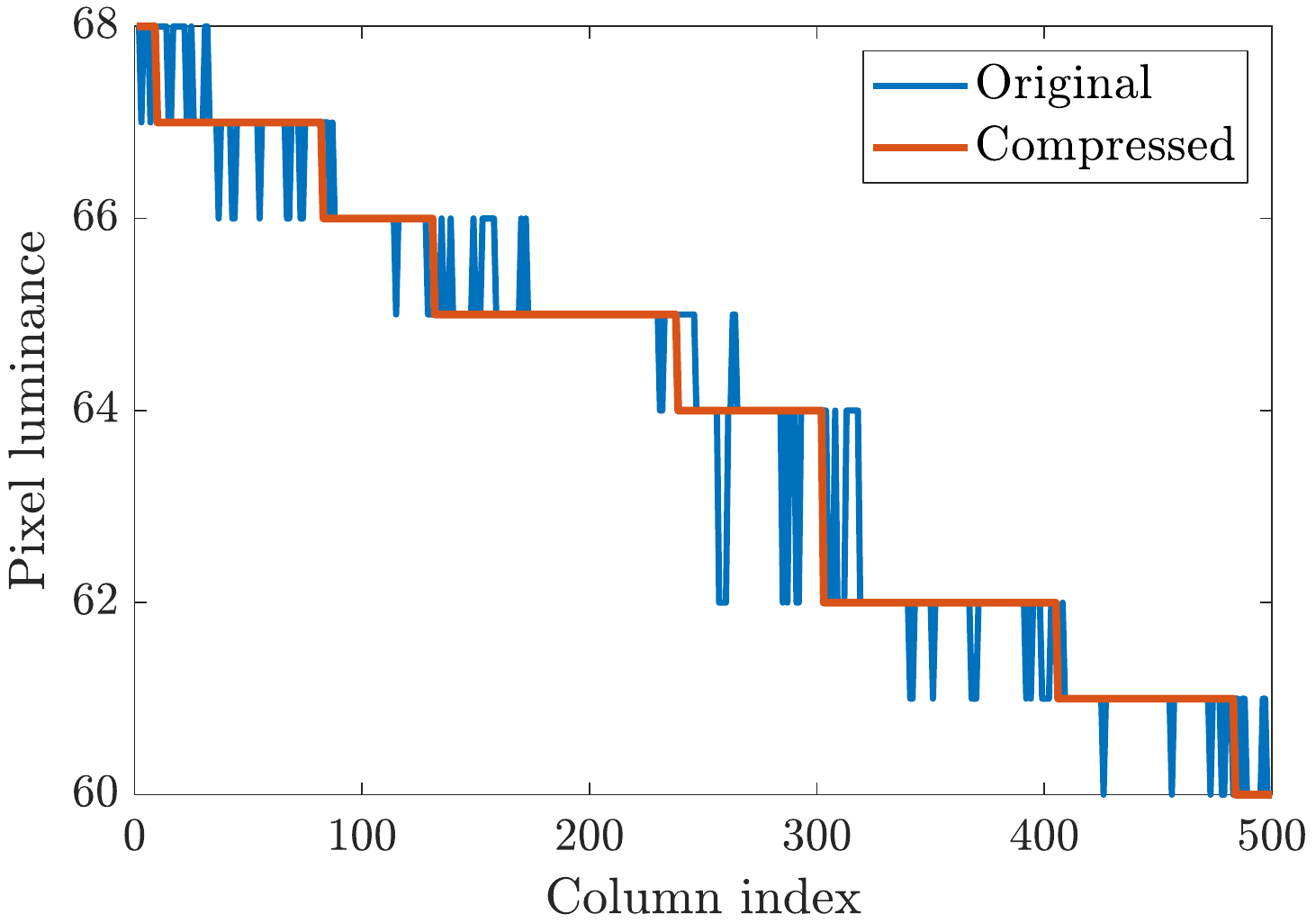}  & 
 \includegraphics[width=\imghei]{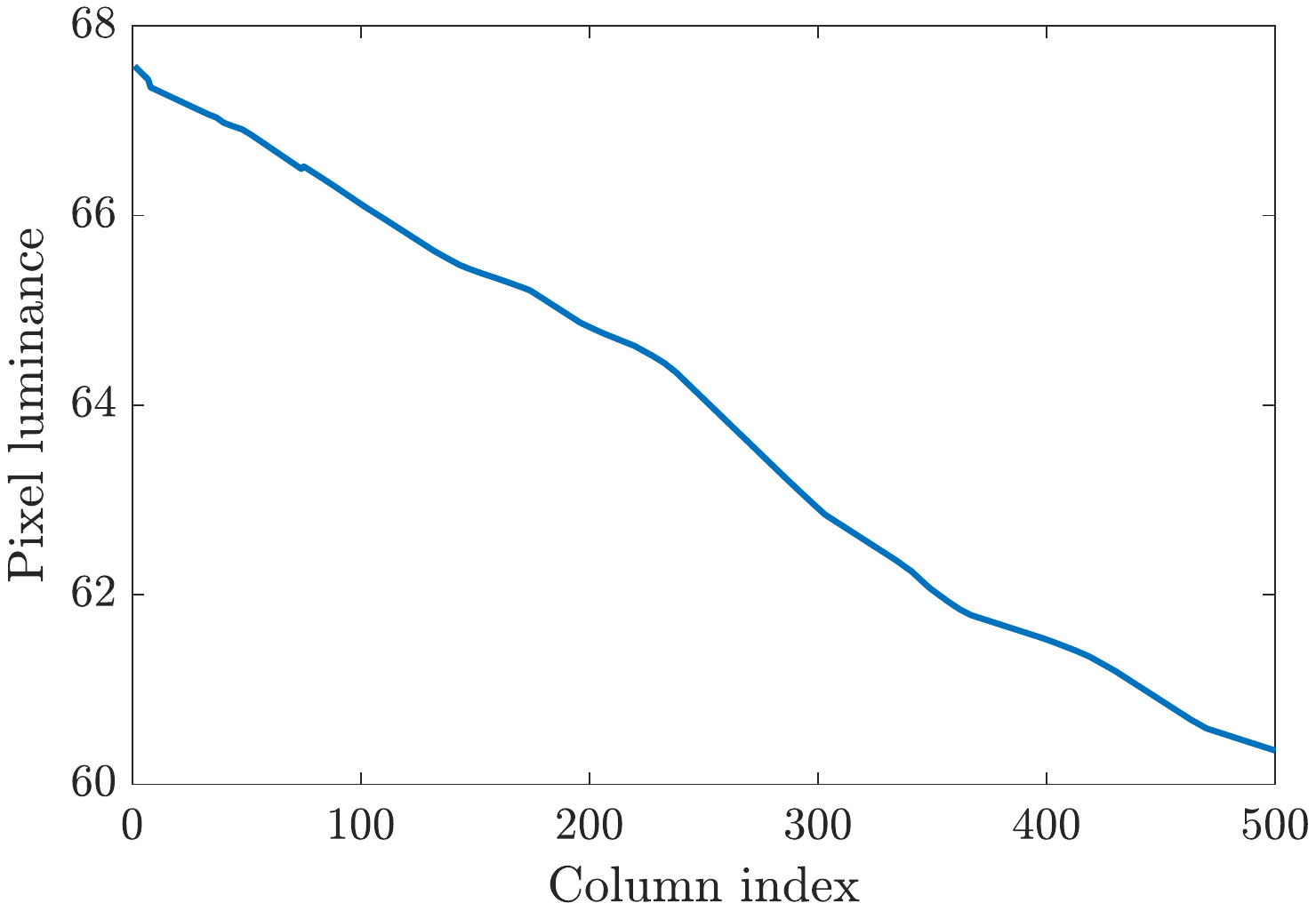}  \\
   (a) Original and compressed  &
   (b) Estimated reconstruction \\
 \includegraphics[width=\imghei]{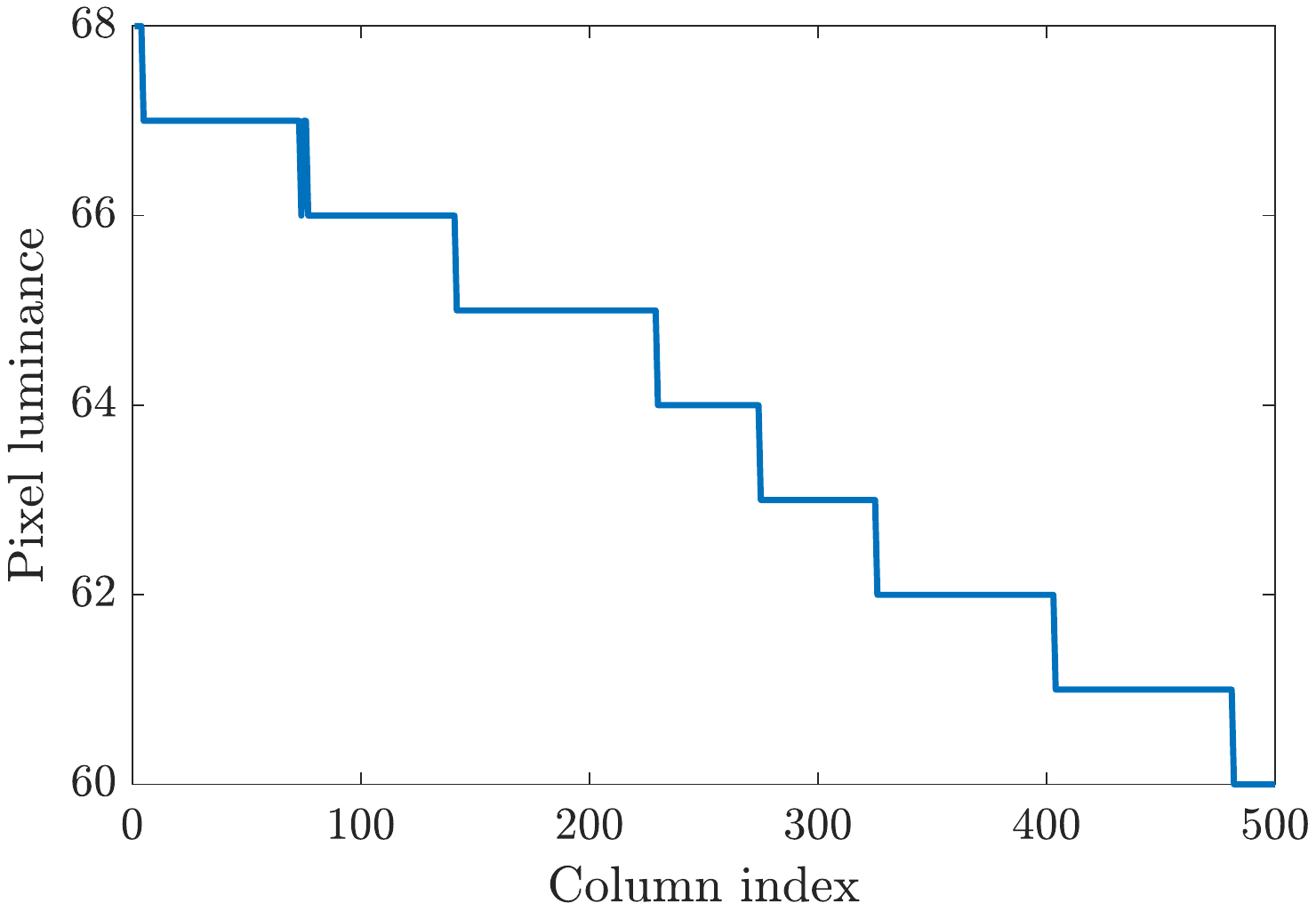}   &
 \includegraphics[width=\imghei]{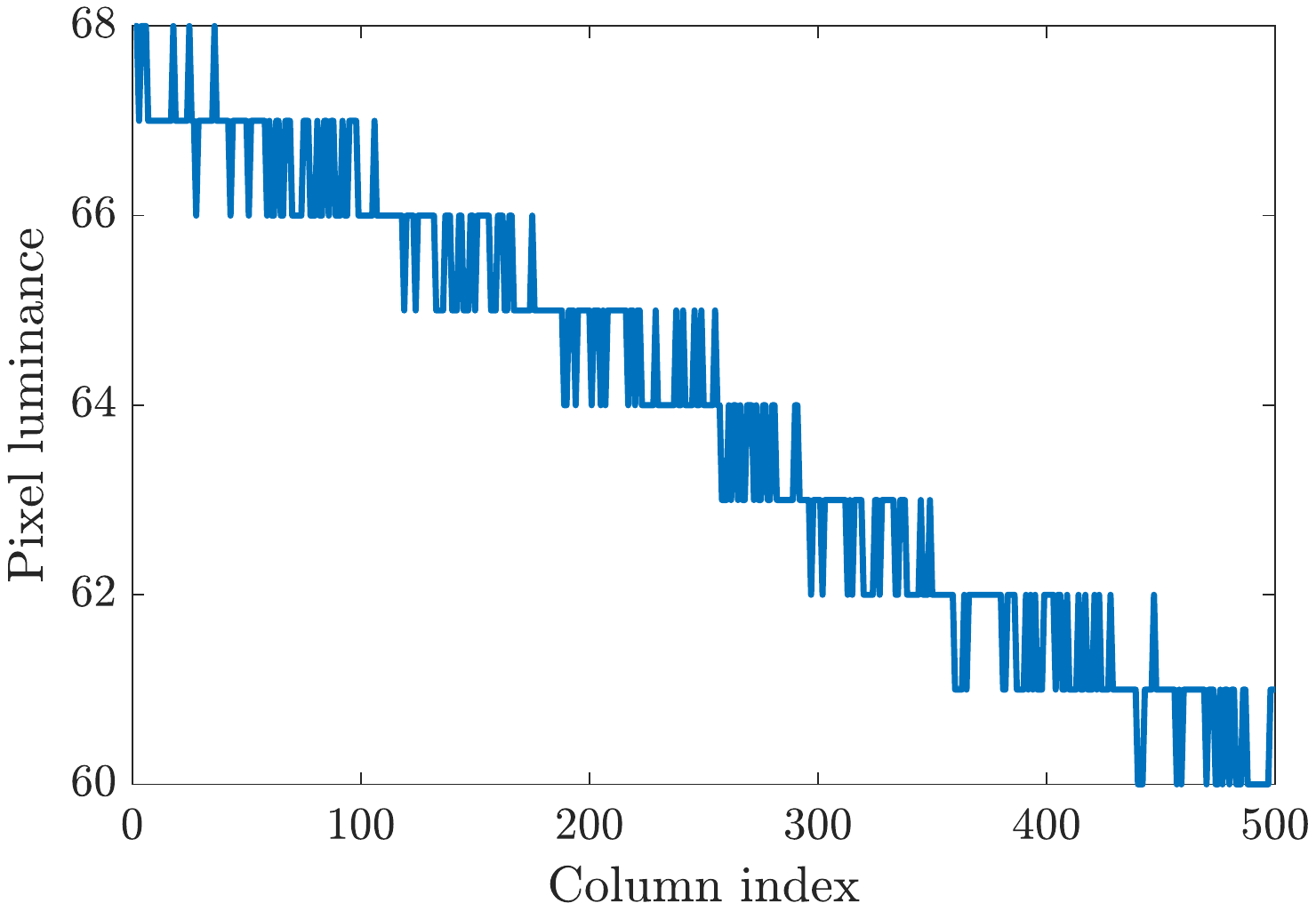}   \\
   (c) Uniform quantization & 
   (d) Dithered quantization \\
\end{tabular}

\caption{One-dimensional visualization of the effects of dithered quantization, using row 200 of the exemplary patch in Fig. \ref{fig:flowchart}. The uniform quantizer on the (b) reconstructed plane still yields banding, as shown in (c), while dithered quantization generates noisy patterns (as in (d)), similar to the original in (a).}
\label{fig:3}
\end{figure}

Fig. \ref{fig:3} shows an example of the effects of processing banded areas with and without dithering, respectively. It may be observed that dithered quantization on the reconstructed banded regions is able to reduce re-quantization error, yielding a pixel distribution similar to the original, while direct quantization without dithering still retains bands. Formally, randomized (dithered) quantization \cite{wannamaker2000theory} may be be expressed as:
\begin{equation}
\label{eq:dither}
\mathcal{P}(i,j)=\boldsymbol{\mathrm{Q}}\big[\mathcal{J}(i,j)+\mathcal{N}(i,j)\big],
\end{equation}
where $\mathcal{J}(i,j)$ is the filtered image calculated via Eq. (\ref{eq:lpf}), $\mathcal{N}(i,j)$ is a 2D noise image, and $\boldsymbol{\mathrm{Q}}[\cdot]$ is a N-to-8-bit quantizer.

\begin{figure*}[!t]
\centering
\def\xlinewidth{0.14}
\def\hswidth{-0.4em}
\def\xem{2pt}
\footnotesize
\setlength{\tabcolsep}{2pt}
\begin{tabular}{cccccc}

  \includegraphics[ width=\xlinewidth\linewidth, keepaspectratio]{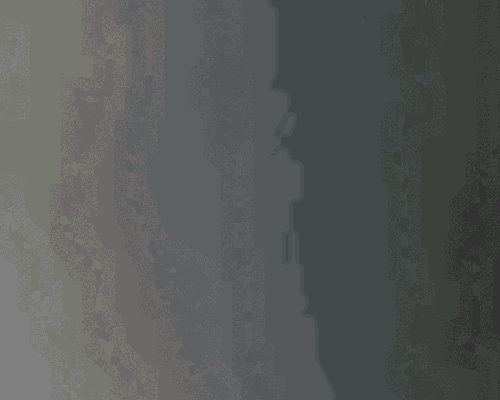} &
  \includegraphics[ width=\xlinewidth\linewidth, keepaspectratio]{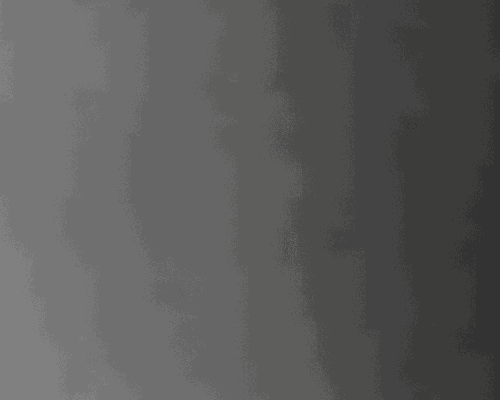} &
  \includegraphics[ width=\xlinewidth\linewidth, keepaspectratio]{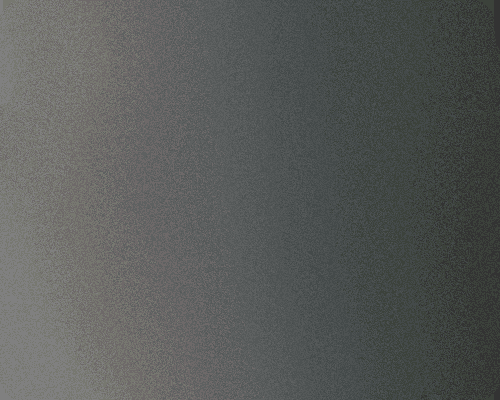} &
  \includegraphics[ width=\xlinewidth\linewidth, keepaspectratio]{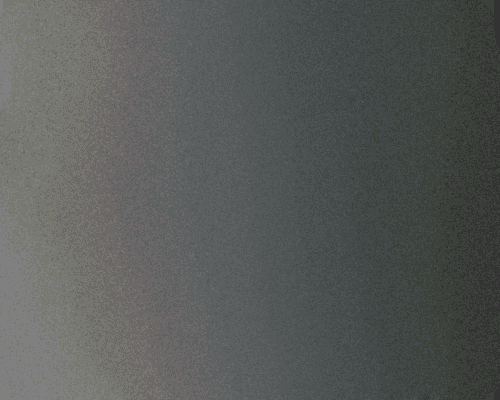} &
  \includegraphics[ width=\xlinewidth\linewidth, keepaspectratio]{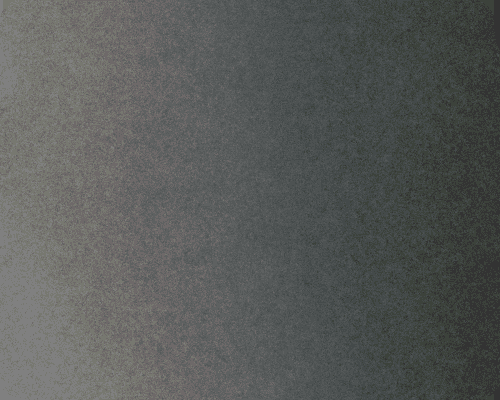} &
  \includegraphics[ width=\xlinewidth\linewidth, keepaspectratio]{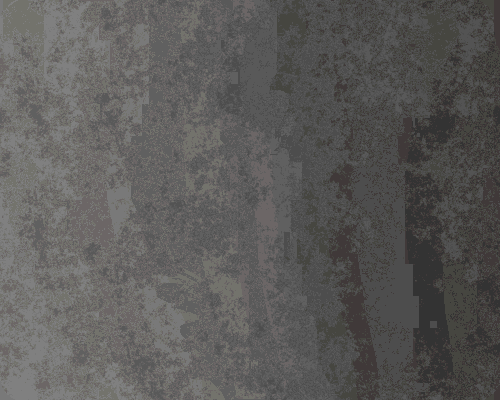}  \\[\xem]
  
  \includegraphics[ width=\xlinewidth\linewidth, keepaspectratio]{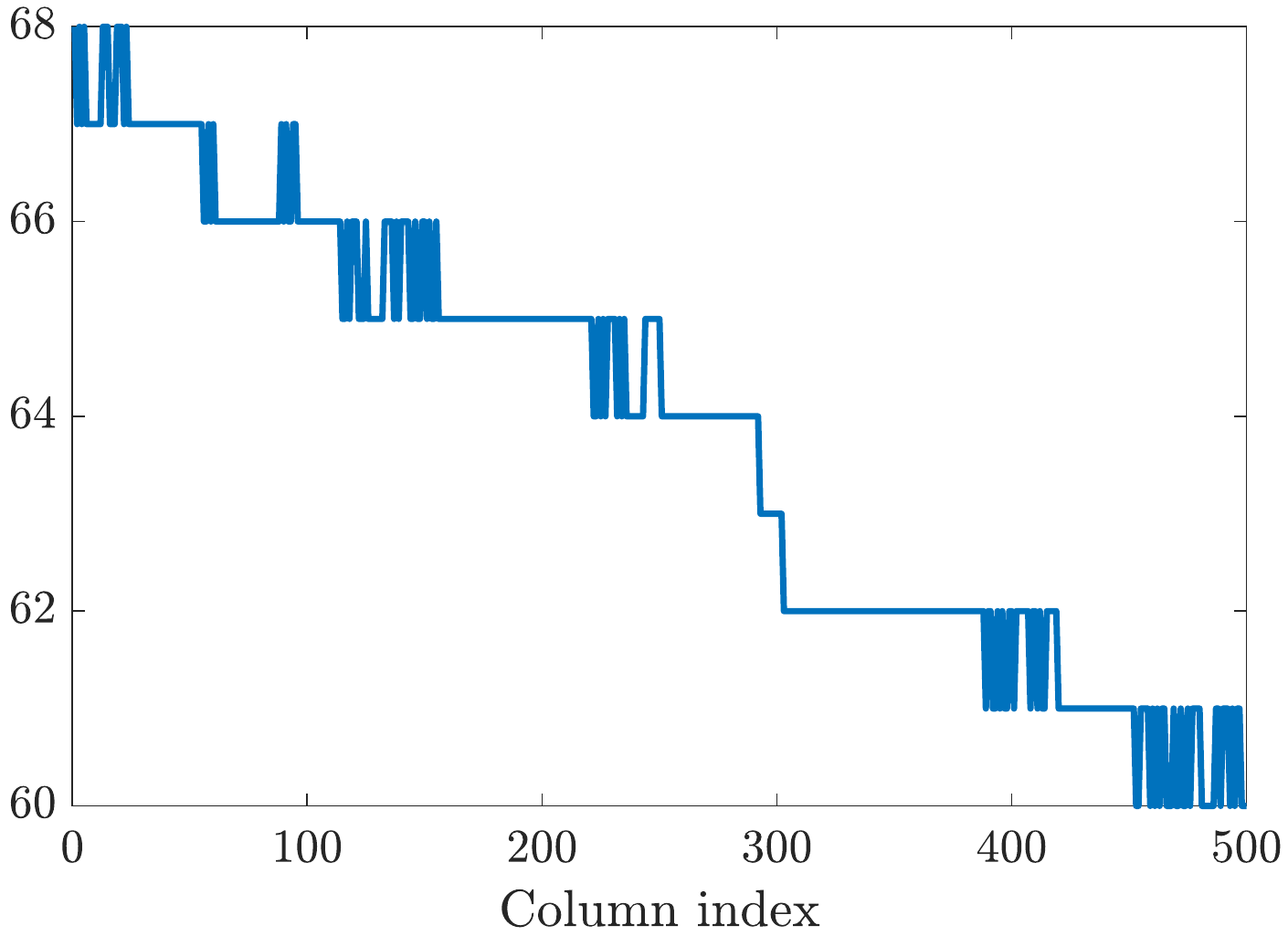} &
  \includegraphics[ width=\xlinewidth\linewidth, keepaspectratio]{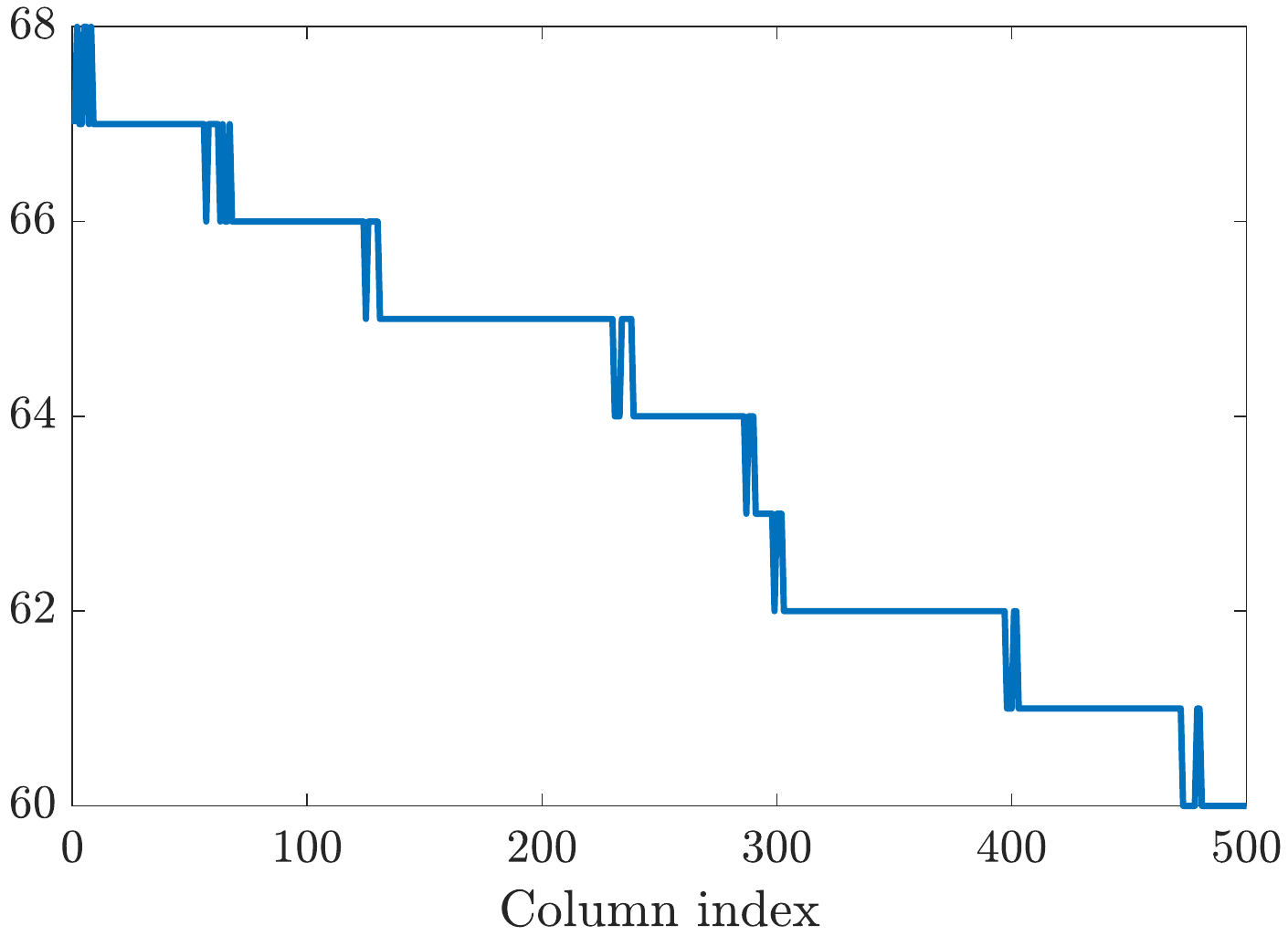} &
  \includegraphics[ width=\xlinewidth\linewidth, keepaspectratio]{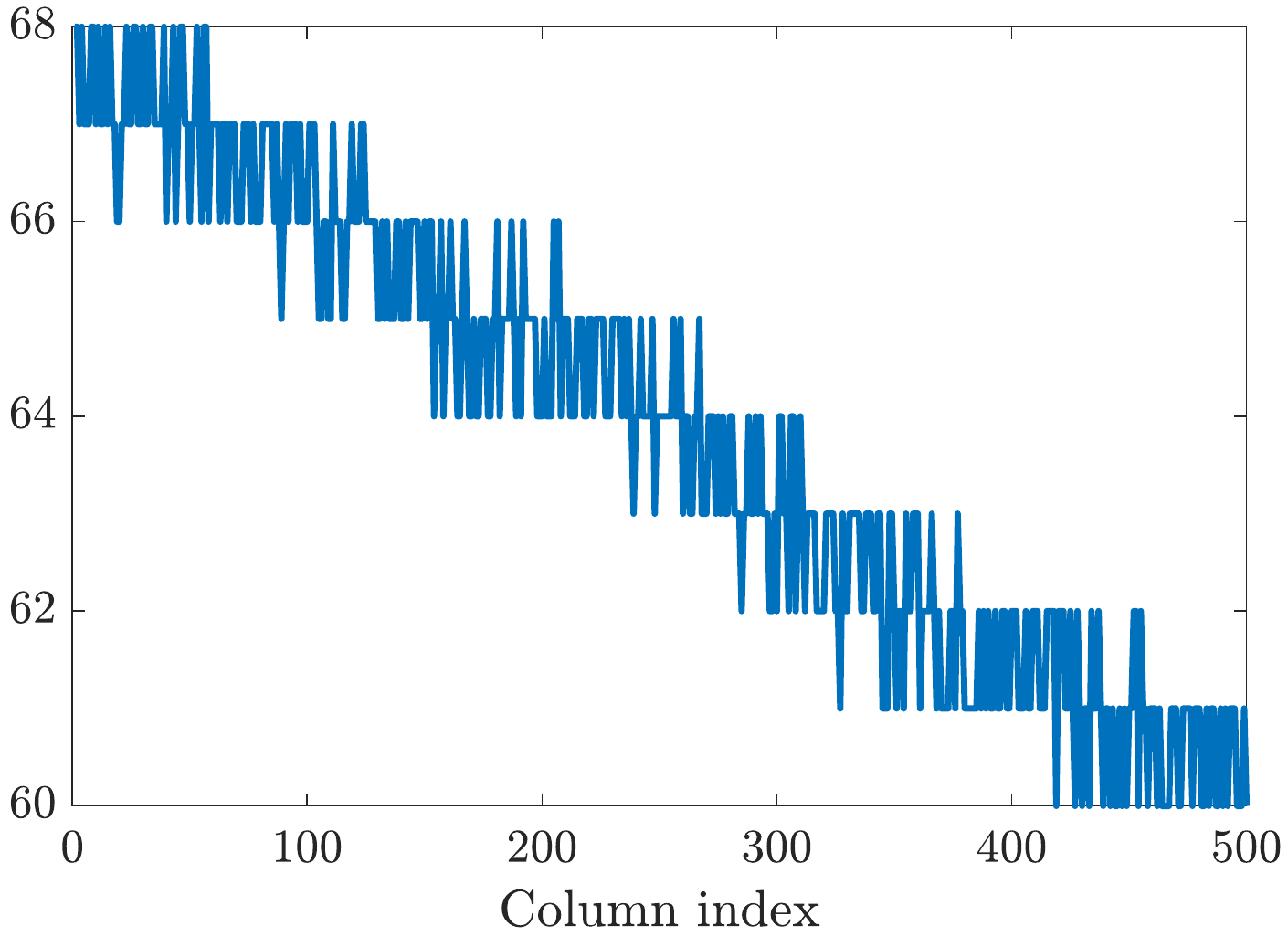} &
  \includegraphics[ width=\xlinewidth\linewidth, keepaspectratio]{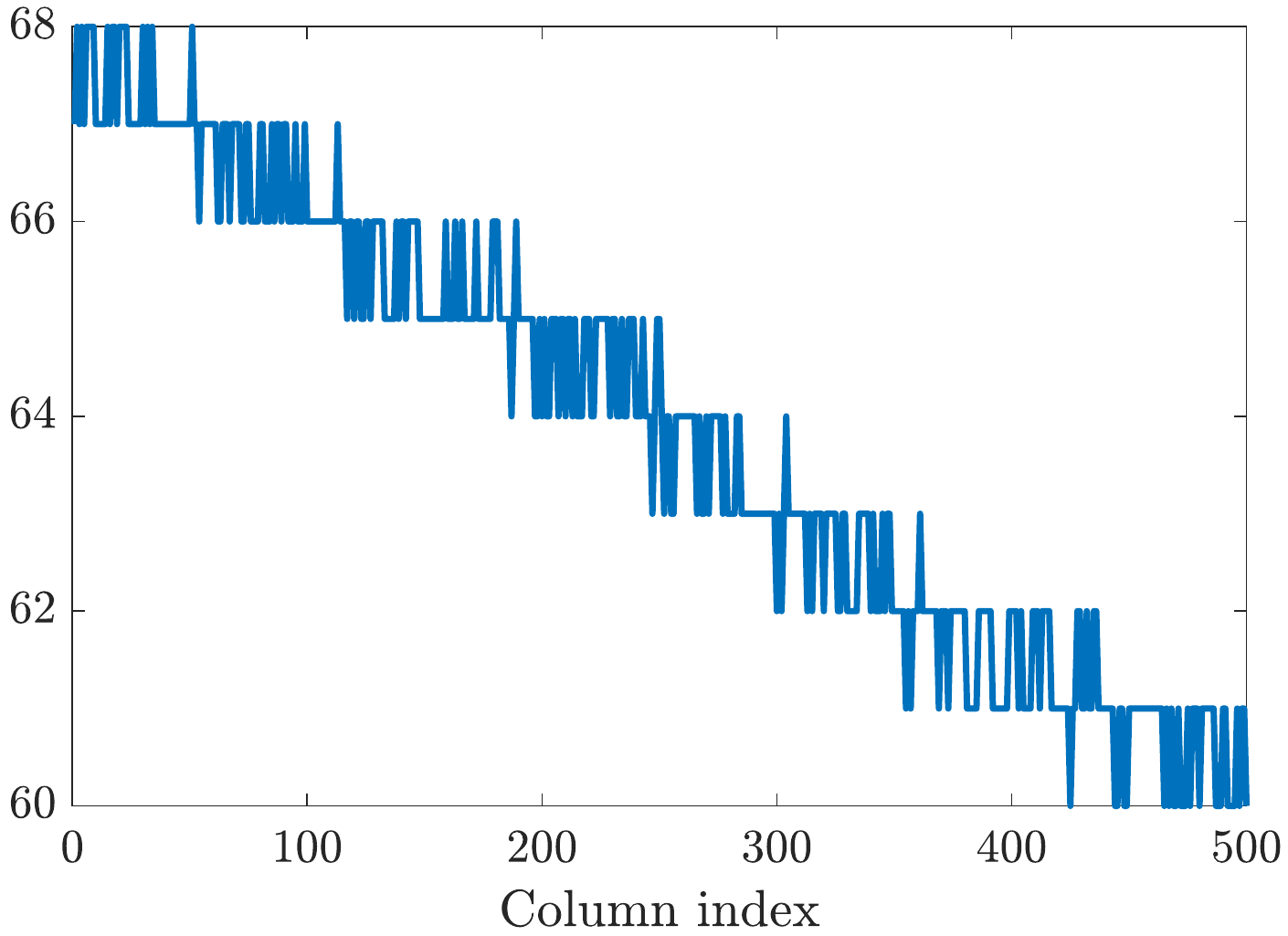} &
  \includegraphics[ width=\xlinewidth\linewidth, keepaspectratio]{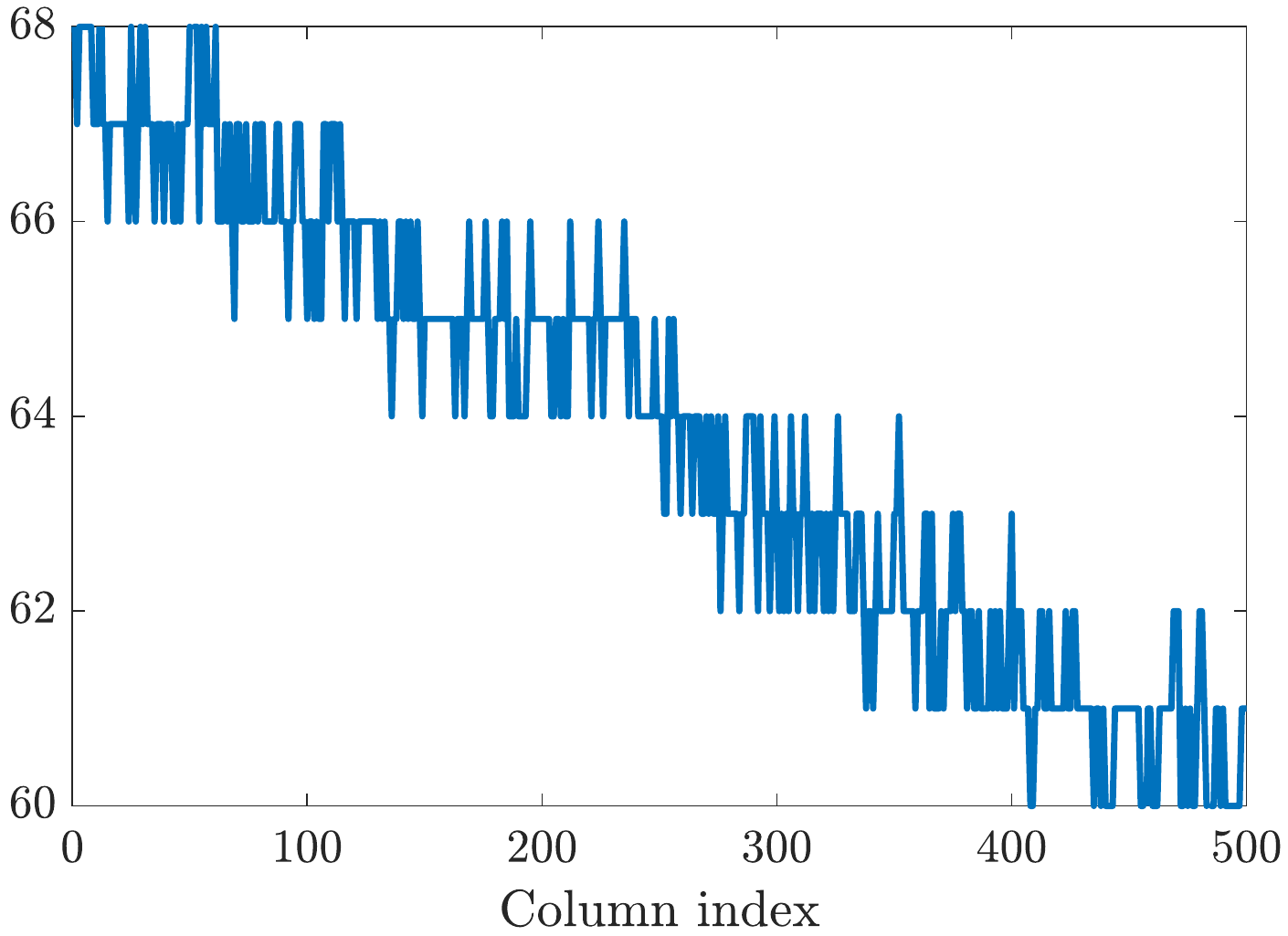} &
  \includegraphics[ width=\xlinewidth\linewidth, keepaspectratio]{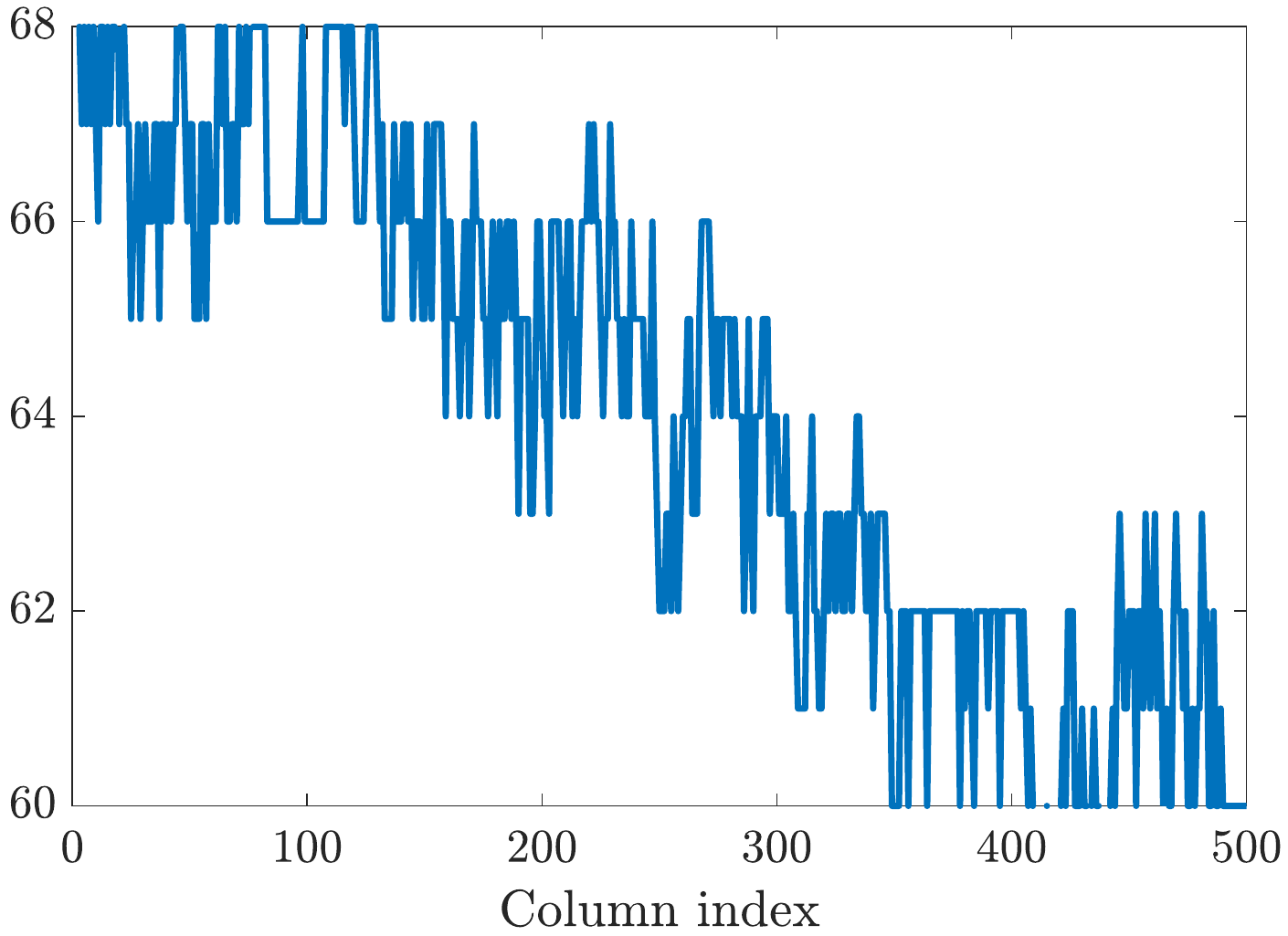}  \\[\xem]
  
 (a) FCDR &
 (b) FFmpeg-deband &
 (c) AdaDeband-UN & 
 (d) AdaDeband-GUN & 
 (e) AdaDeband-$f^{-0.5}$ &
 (f) AdaDeband-$f^{-1}$ \\
 
\end{tabular}
\caption{Visualizations (1st row) and corresponding pixel value distributions (2nd row) of compared debanding outputs on the exemplary patch in Fig. \ref{fig:flowchart}. The compared methods are (a) FCDR \cite{huang2016understanding}, (b) FFmpeg-deband \cite{ffmpeg-deband}, (c-f) AdaDeband with (c) uniform noise (UN), (d) Gaussian-blurred uniform noise (GUN), (e) $1/f^{1/2}$ noise, and (f) pink noise ($1/f$). The top row of figures has been contrast-stretched for better visualization.}
\label{fig:noise_pattern}
\end{figure*}

\begin{figure*}[!ht]
	\centering
	\footnotesize
	\renewcommand{\tabcolsep}{3pt} 
	\renewcommand{\arraystretch}{1} 
	\def\imgwid{0.27\textwidth}
	\def\rdhei{0.223\textwidth}
	\def\imghei{0.18\textwidth}
	\def\imgheid2{0.088\textwidth}
	
	\def\rd_shift{-0.123\textwidth}
    \def\im_shift{-0.093\textwidth}
	
	\begin{tabular}{cccccc}
	
        Source Video &
        Original &
        Compressed &
        FCDR &
        FFmpeg &
        AdaDeband
        \\ \\[-2ex]
        \multirow[t]{2}{*}[\im_shift]{\includegraphics[height=\imghei]{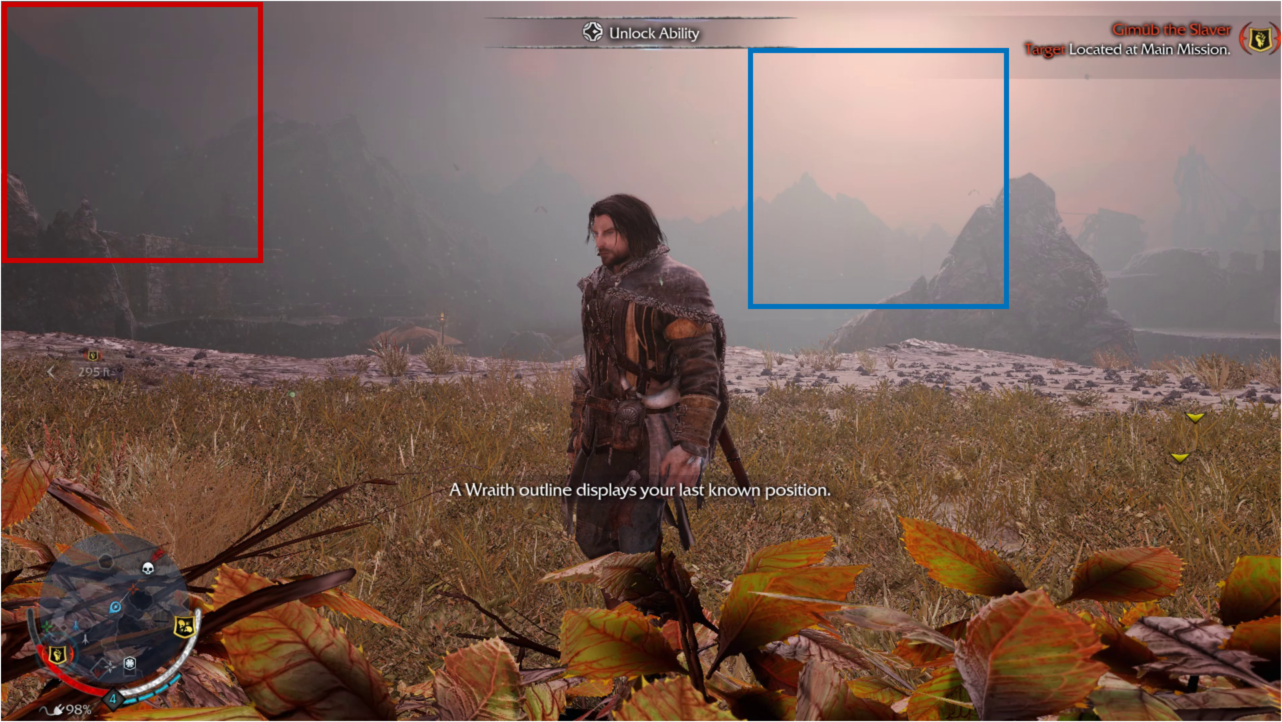}} &
        \includegraphics[height=\imgheid2]{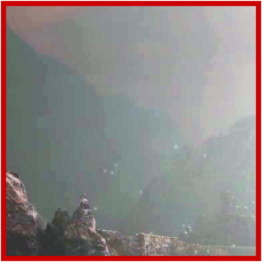} & 
        \includegraphics[height=\imgheid2]{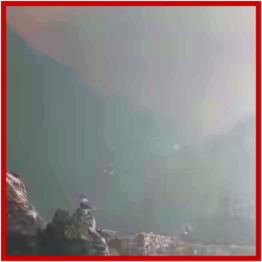} & 
        \includegraphics[height=\imgheid2]{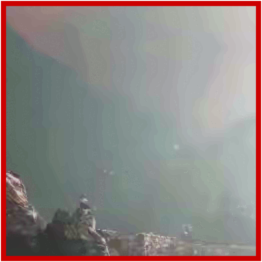} &
        \includegraphics[height=\imgheid2]{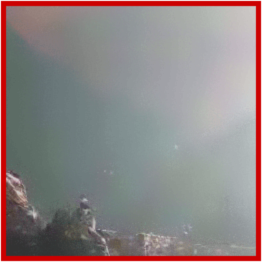} & 
        \includegraphics[height=\imgheid2]{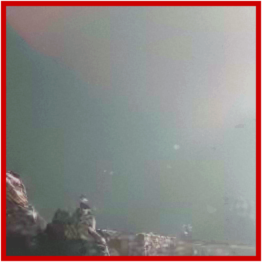} \\
		&  
		\includegraphics[height=\imgheid2]{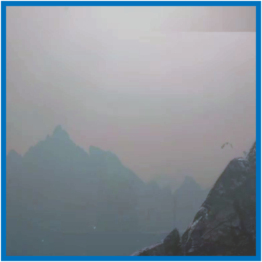} &
        \includegraphics[height=\imgheid2]{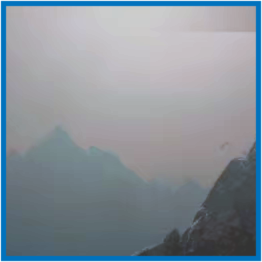} &
        \includegraphics[height=\imgheid2]{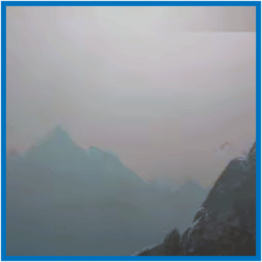} &
        \includegraphics[height=\imgheid2]{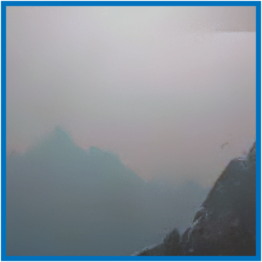} &
        \includegraphics[height=\imgheid2]{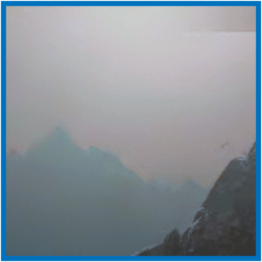} \\
        
          \multirow{2}{*}{(a) {Gaming\_1080P-71a5}}
         & PSNR / SSIM 
         & 30.49 / 0.888 
         & 30.49 / 0.888 
         & 30.46 / 0.885  
         & 30.48 / 0.887 \\
         & BBAND 
         & 0.670 
         & 0.724
         & 0.261 
         & 0.250\\
        \multirow[t]{2}{*}[\im_shift]{\includegraphics[height=\imghei]{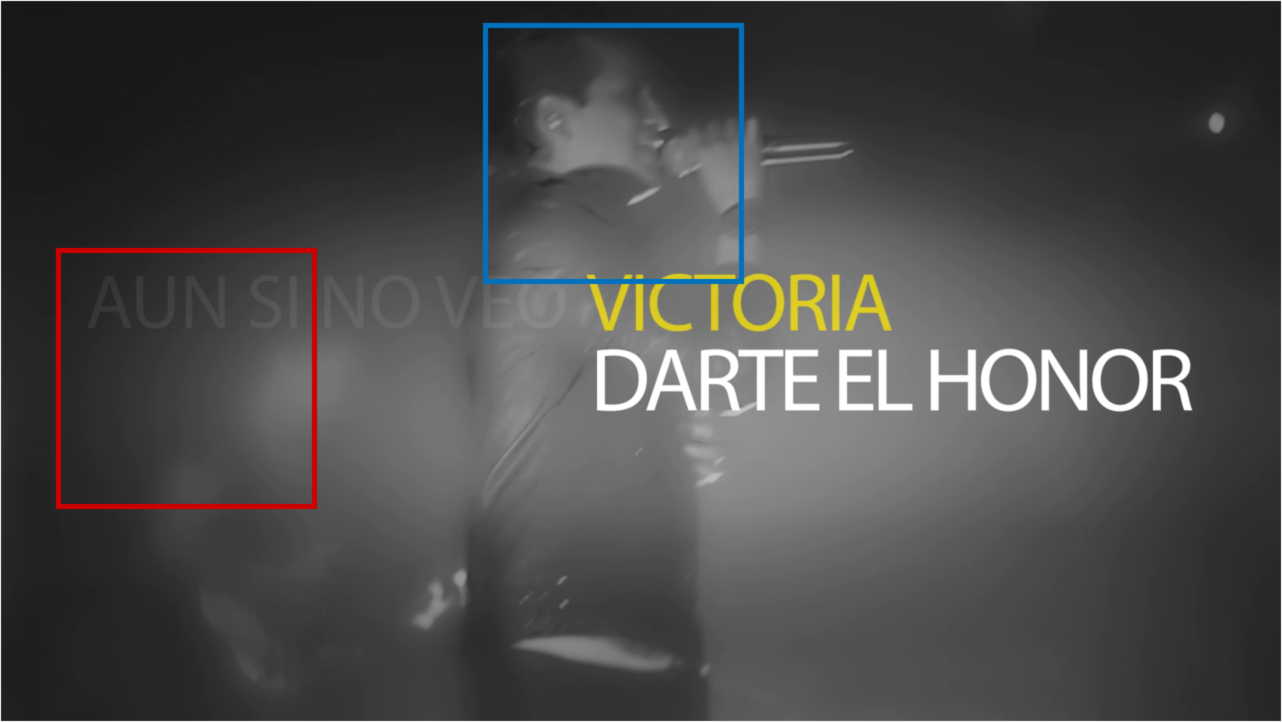}} &
        \includegraphics[height=\imgheid2]{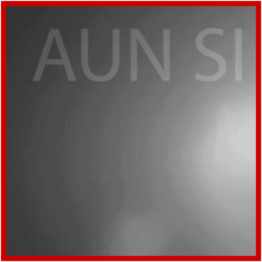} & 
        \includegraphics[height=\imgheid2]{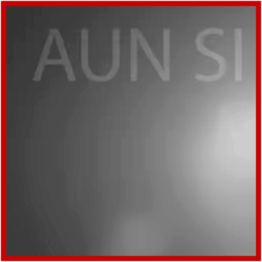} & 
        \includegraphics[height=\imgheid2]{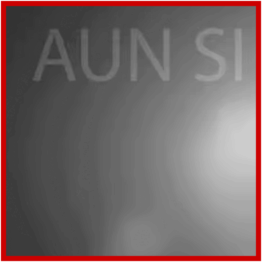} &
        \includegraphics[height=\imgheid2]{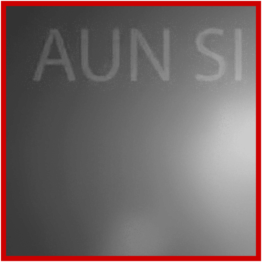} & 
        \includegraphics[height=\imgheid2]{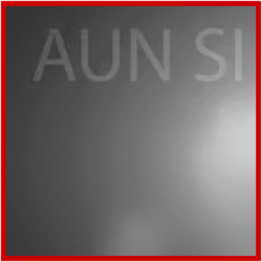} \\
		&  
		\includegraphics[height=\imgheid2]{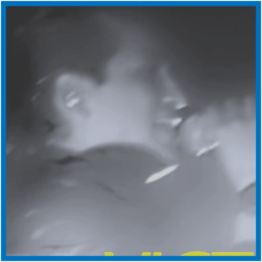} &
        \includegraphics[height=\imgheid2]{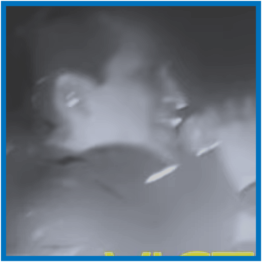} &
        \includegraphics[height=\imgheid2]{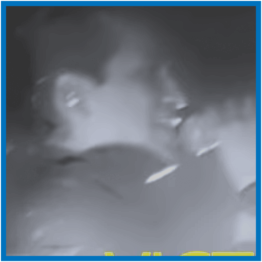} &
        \includegraphics[height=\imgheid2]{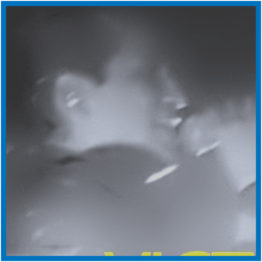} &
        \includegraphics[height=\imgheid2]{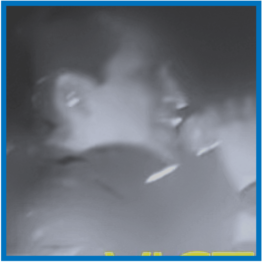} \\
        
          \multirow{2}{*}{(b) {LyricVideo\_1080P-5a1f}}
         & PSNR / SSIM 
         & 45.12 / 0.994 
         & 44.93 / 0.993 
         & 44.45 / 0.990 
         & 44.35 / 0.991 \\
         & BBAND 
         & 1.274 
         & 0.914
         & 0.377
         & 0.363  \\
	\end{tabular}
	\caption{Qualitative and quantitative comparisons of debanded frames produced by different debanding algorithms. Frame crops (contrast-stretched for better visualization) from left to right: original, compressed, FCDR, FFmpeg-deband, and AdaDeband. We follow the style in \cite{chen2019proxiqa}.}
	\label{fig:visual_comp}
\end{figure*}

It should be noted that the pattern of the noise image $\mathcal{N}(i,j)$ used for dithering will shape the textures of the resulting debanded regions, hence the type of noise must be carefully selected. We demonstrate several well-known noise patterns and their corresponding dithered outcomes, both visually and quantitatively, in Fig. \ref{fig:noise_pattern}. Among commonly used methods \cite{ulichney1988dithering, lippel1971effect, ulichney1999review}, we chose to employ Gaussian blurred uniform white noise $\mathcal{U}(-2,+2)$, as shown in Fig. \ref{fig:noise_pattern}(d); other recommended and effective options include uniform and $1/f^{1/2}$ noise (Fig. \ref{fig:noise_pattern}(c)(e)). It may also be observed from Fig. \ref{fig:noise_pattern} that AdaDeband outperforms FCDR and FFmpeg-deband when smoothing the banded staircase, yielding a more pleasant visual enhancement.

\renewcommand{\footnotesize}{\fontsize{7pt}{8pt}\selectfont}
\section{Experiments}
\label{sec:exp}

We compared our proposed adaptive debanding filter (dubbed AdaDeband) against two recent debanding/decontouring methods proposed for compressed videos: FCDR \cite{huang2016understanding} and the FFmpeg-deband filter \cite{ffmpeg-deband}, on ten\footnote{Filenames: Gaming\_1080P-71a5, NewsClip\_1080P-2eb0, Sports\_1080P-19d8, Vlog\_720P-60f8, LyricVideo\_1080P-3b60, LyricVideo\_1080P-5a1f, MusicVideo\_720P-44c1, MusicVideo\_720P-3698, Sports\_720P-058f, Vlog\_720P-32b2\label{fn1}} selected videos from the YouTube UGC dataset \cite{wang2019youtube}. All the test sequences were scaled to 720p for computational convenience, and we compressed the videos using VP9 constrained quality (CQ) mode with \texttt{-crf 39} to generate noticeable banding artifacts. In our implementations, only the luma channel was filtered since much less banding was observed on the Cb/Cr channels, but the proposed AdaDeband could also be applied to each color or chroma component.

Visual comparisons of the debanding methods are shown in Fig. \ref{fig:visual_comp}. We may see that AdaDeband effectively smoothed the banding, leaving edges/textures well preserved. FFmpeg-deband, in contrast, tended to over-smooth weak textures, and under-smooth relatively large banded regions. Another advantage of AdaDeband is its adaptiveness, which can remove bands of any scale/shape, whereas FFmpeg-deband and FCDR require specification of a set of filter parameters, which may affect performance on scenarios it has not been exposed to.

\begin{figure}[!t]
\centering
\includegraphics[width=0.26\textwidth]{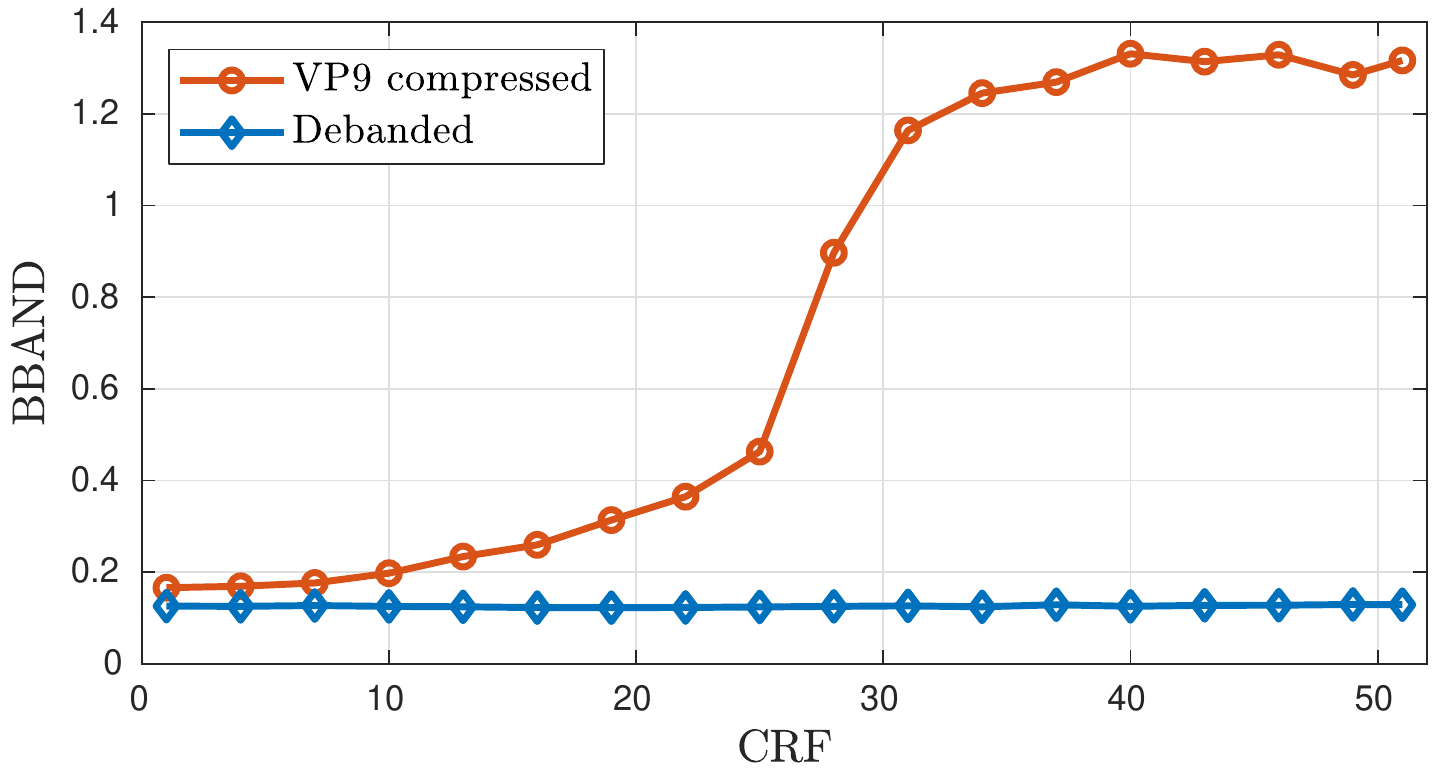}
\caption{AdaDeband is robust against different levels of quantization.}
\label{fig:crf}
\end{figure}

To further verify the adaptiveness of AdaDeband against different quantization parameters, we generated different levels of banding effects using the VP9 CQ mode, with \texttt{-crf} ranging from 1 to 51 on the exemplary video in Fig. \ref{fig:example_deband}. We may see in Fig. \ref{fig:crf} that our proposed debanding filter produced robust BBAND scores against different levels of quantization without any further tuning of filter parameters.

In addition to the visual results, we also quantitatively compared the debanded videos using several common video quality models. However, it has been shown that traditional video quality metrics like PSNR and even SSIM family \cite{wang2004image, pei2015image, wang2003multiscale} do not align very well with human perception of banding \cite{wang2016perceptual}. Moreover, if the original video already contains banding artifacts, which is often the case for high-resolution videos, it is less reliable to rely on full-reference quality models, since ``more similarity'' with respect to the original is not necessarily indicating ``perceptually better.'' Therefore, we also compare on a blind banding assessment model (the BBAND index) \cite{tu2020bband}, which has been shown to deliver predictions consistent with subjective judgments of banding. It may be seen in Table \ref{table:perf} that AdaDeband outperformed FCDR and is on par with FFmpeg-deband in terms of BBAND scores, indicating that it produces perceptually favorable results. Moreover, AdaDeband yields slightly better PSNR and SSIM scores than FFmpeg-deband, indicating less distortion compared to the original.

\begin{table}
\renewcommand{\arraystretch}{1}
\setlength{\tabcolsep}{5pt}
\caption{Performance comparison of debanding methods. Each cell shows the evaluation results formatted as MEAN ($\pm$STD).}
\label{table:perf}
\centering
\begin{tabular}{l|cccc}
\hline\hline
Method &  PSNR$\uparrow$ &  SSIM$\uparrow$ &  BBAND$\downarrow$ \\
\hline
FCDR & 39.08 ($\pm$5.10) & 0.9709 ($\pm$0.0308) & 0.5790 ($\pm$0.2494) \\
FFmpeg & 38.84 ($\pm$4.96) & 0.9677 ($\pm$0.0311) & 0.2264 ($\pm$0.0903) \\
AdaDeband & 38.97 ($\pm$4.97) & 0.9699 ($\pm$0.0309) & 0.2206 ($\pm$0.0895) \\
\hline\hline
\end{tabular}
\end{table}

\section{Conclusion}
\label{sec:conc}

We proposed an adaptive post-debanding filter to remove banding artifacts resulting from coarse video compression. The efficacy of the algorithm can be attributed to the accurate detection of banding regions, content-aware band reconstruction, and a dithered re-quantization. Both visual and quantitative results demonstrate significant performance improvement over prior debanding methods. Actually, it can be regarded as a visual enhancement algorithm, whose effects could be accounted for when performing rate-distortion optimized video encoding. Since it is a post-processing model, it may be optimized for efficient, low-power implementations in real-time applications.

\bibliographystyle{IEEEtran}
\bibliography{refs}

\end{document}